\documentclass[aps,prl,twocolumn,superscriptaddress,showpacs]{revtex4}

\usepackage{graphicx}

\begin{document}


\title{Static and Dynamic Magnetism in Underdoped  Superconductor BaFe$_{1.92}$Co$_{0.08}$As$_2$}


\author{A. D. Christianson}
\author{M. D. Lumsden}
\author{S. E. Nagler}
\author{G. J. MacDougall}
\author{M. A. McGuire}
\author{A. S. Sefat}
\author{R. Jin}
\author{B. C. Sales}
\author{D. Mandrus}
\affiliation{Oak Ridge National Laboratory, Oak Ridge, TN 37831, USA}

\begin{abstract}

We report neutron scattering measurements on single crystals of BaFe$_{1.92}$Co$_{0.08}$As$_2$.  The magnetic Bragg peak intensity is reduced by 6 $\%$ upon cooling through T$_C$.  The spin dynamics exhibit a gap of 8 meV with anisotropic three-dimensional (3d) interactions.  Below T$_C$ additional intensity appears at an energy of  $\sim$4.5(0.5) meV similar to previous observations of a spin resonance in other Fe-based superconductors.  No further gapping of the spin excitations is observed below T$_C$ for energies down to 2 meV.  These observations suggest the redistribution of spectral weight from the magnetic Bragg position to a spin resonance demonstrating the direct competition between static magnetic order and superconductivity.

\end{abstract}


\maketitle

Despite intense experimental and theoretical efforts directed towards understanding the recently discovered Fe-based superconductors\cite{LaOFFeAsdis,SmOFFeAsdis,CeOFFeAsdis,HPdis,Rotter,Wang,Yeh}, the mechanism behind superconductivity remains unclear (\textit{e.g.} \cite{singhlda,cvetkovic,bang,chen_theory,chubukov_review, mazin_review}).  However, one common feature is the presence of magnetism in large regions of the phase diagrams for these materials (\textit{e.g.} \cite{ cruz_lafeaso,mcquire_big,Huang,lynn_review}).  As such, there have been many suggestions that the pairing mechanism originates from the spin fluctuations.  The observation of a resonance, strongly coupled to T$_C$, in the spin dynamics of BaFe$_2$As$_2$ doped with potassium\cite{christianson2}, cobalt\cite{lumsden_co}, and nickel\cite{chi_ni} supports this contention and emphasizes the need for a comprehensive understanding of the fundamental magnetic behavior of these materials.

Consequently, numerous studies have been initiated to probe the magnetic behavior of the Fe-based superconductors.  In AFe$_2$As$_2$ (A=Ba,Sr,Ca) studies of the static magnetic properties have revealed spin density wave order with magnetic moments aligned along the a-axis of the low temperature orthorhombic structure\cite{lynn_review}.  The spin excitations of these compounds are the subject of intense scrutiny as means of establishing the degree to which the magnetic behavior originates with localized or itinerant Fe 3\textit{d}-electrons\cite{lynn_review}.  As noted above the spin dynamics in superconducting samples have been studied in several members of optimally doped BaFe$_2$As$_2$ \cite{christianson2,lumsden_co,chi_ni} where both the tetragonal-orthorhombic structural distortion and long range magnetic order have been suppressed.  In particular, for doping levels where static magnetism coexists with superconductivity (either microscopically or as separate phases) the nature of the magnetic excitations has yet to be established in detail.

BaFe$_{2-x}$Co$_x$As$_2$\cite{Sefat3} has emerged as one of the most important systems in large part due to the availability of large homogenous single crystals. The phase diagram \cite{Ning,chu_cophase,pratt} for these materials shows that in the underdoped regime, there is both a structural phase transition and a magnetic ordering transition, as in the parent compound.  For the range of concentrations, 0.06 $< x <$ 0.12, samples exhibit both magnetic order and superconductivity.  The topic of this letter is a neutron scattering study of single crystal samples of underdoped BaFe$_{1.92}$Co$_{0.08}$As$_2$  with T$_N$=58 K and $T_C$=11 K.  We observe a 6 $\%$ reduction of the magnetic Bragg peak intensity on cooling through T$_C$.  The reduction in intensity is not due to a change in the magnetic structure and hence represents either a reduction in the ordered moment or a decrease in the fraction of the sample exhibiting long range magnetic order.  On the other hand, the spin dynamics associated with the magnetic ordering is gapped above T$_C$ while below T$_C$ additional intensity appears at an energy of  $\sim$4.5(0.5) meV consistent with the observation of a spin resonance in other Fe-based superconductors.  Thus, a loss in spectral weight in the magnetic Bragg peaks appears to be compensated by the appearance of a spin resonance highlighting the competition between magnetic order and superconductivity.  The intensity of the resonance observed for x=0.08 depends on \textbf{Q} both in-plane and along the c-axis in stark contrast to the two-dimensional resonance observed in the optimally doped sample \cite{lumsden_co}.  We note that the neutron scattering data presented here does not allow for differentiation between microscopic coexistence and phase separation of superconductivity and magnetism.

\begin{figure}
\centering\includegraphics[width=1.0\columnwidth]{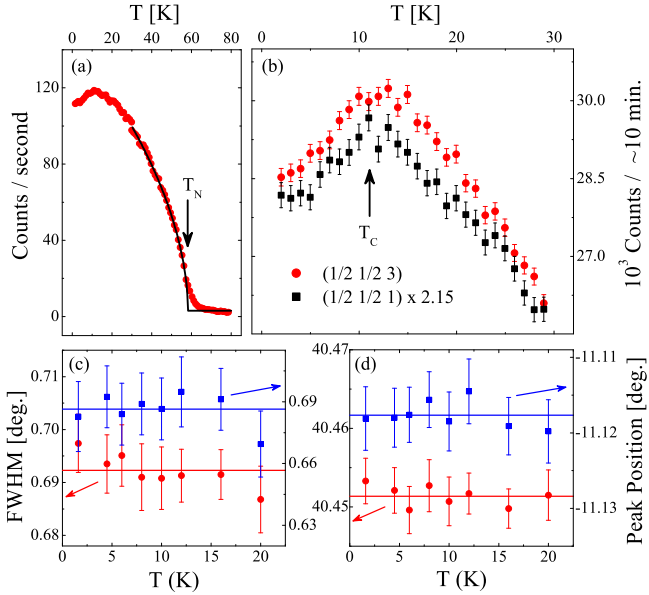}
\caption{\label{fig1} (a) Peak intensity of the (1/2 1/2 3) magnetic Bragg peak.  The solid line is the result of a power law fit as described in the text.  (b) Peak intensity of the (1/2 1/2 1) (squares) and (1/2 1/2 3) (circles) magnetic peaks through T$_C$.  The FWHM (c) and peak position (d) extracted from transverse and longitudinal scans in terms of $\theta$ (squares) and ${2\theta}$ (circles) for the (1/2 1/2 3) peak through T$_C$. The solid lines in the inset of (b) and in (c) and (d) represent the average of the data weighted by the statistical errors.}
\end{figure}

Single crystals of BaFe$_{1.92}$Co$_{0.08}$As$_2$ were grown using FeAs flux as described in \cite{lumsden_co}.  Magnetic susceptibility and specific heat measurements show that T$_C$=11 K.  Although at low temperature the crystal structure is strictly orthorhombic\cite{chu_cophase,pratt}, the measurements reported here have insufficient resolution to observe the consequences of the structural distortion.  For this reason and for consistency with previous inelastic measurements on superconducting samples \cite{christianson2,lumsden_co} we use tetragonal notation exclusively.  For reference (1 0 L), the antiferromagnetic ordering wave vector in orthorhombic notation, is indexed as (1/2 1/2 L) in tetragonal notation.  Further discussion of the various notations employed in the literature may be found in \cite{lumsden_co,mazin}.

Neutron diffraction data was collected on a 0.2 g crystal.  For the inelastic neutron scattering measurements, four single crystals of BaFe$_{1.92}$Co$_{0.08}$As$_2$ with a total mass of 2 g were co-aligned in the (HHL) plane.  The data in Fig. 1(b) was collected using the HB1A triple-axis spectrometer (TAS) at the HFIR configured with collimations of 48$^\prime$-40$^\prime$-40$^\prime$-136$^\prime$.  The remainder of the data presented here was collected with the HB-3 TAS at the HFIR configured with collimations of 48$^\prime$-60$^\prime$-80$^\prime$-120$^\prime$ with a fixed final energy of 14.7 meV using pyrolitic graphite monochromator and analyzer crystals. Preliminary inelastic neutron scattering data (not shown) was collected on the HB1 TAS at the HFIR.

\begin{figure}
\centering\includegraphics[width=1.0\columnwidth]{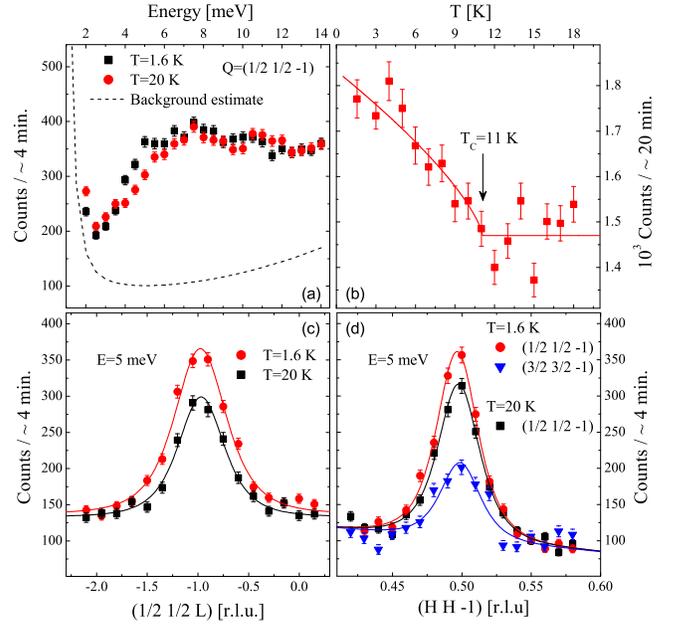}
\caption{\label{fig2} (a) Constant-Q scan, Q=(1/2 1/2 -1), for temperatures of 1.6 K ($T<T_C$) and 20 K ($T>T_C$).  For comparison the background scattering determined as described in text is also shown.  (b)The temperature dependence of the inelastic intensity at (1/2 1/2 -1) and E=5 meV.  The solid line is a power law fit which results in a $T_C$ of 11(1) K consistent with bulk measurements.  L-dependence (c) and H-dependence (d) of the inelastic intensity near Q=(1/2 1/2 -1) and E=5 meV at 1.6 and 20 K.  The solid lines are guides to the eye.  In (d) a scan around (3/2 3/2 -1) is displayed to emphasize the magnetic origin of the scattering.}
\end{figure}

Fig. 1 shows neutron diffraction data as a function of temperature for BaFe$_{1.92}$Co$_{0.08}$As$_2$.  Fig. 1(a) Shows the temperature dependence of the (1/2 1/2 3) magnetic Bragg peak.  Fits to a power law over a limited range of temperatures (30 K $<$ T $<$80 K) yield an ordering temperature, T$_N$=58(0.6) K.
Figure 1(b) indicates a significant reduction in intensity below T$_C$ for both (1/2 1/2 1) and (1/2 1/2 3).  The intensity at 1.6 K is reduced by 6$\%$ relative to that of the maximum intensity found at T$_C$.  The change in intensity of both peaks occurs in the same manner so that the effect cannot be explained by change in magnetic structure or a spin reorientation. Transverse and longitudinal scans were made through (1/2 1/2 1) and (1/2 1/2 3) to ensure that that the observed change is due to a reduction in the integrated intensity rather than a change in peak position or line width.   The full width at half maximum (FWHM) and peak positions extracted from fits to the (1/2 1/2 3) peak are shown in fig. 1(c) and (d). A Gaussian lineshape was used for the longitudinal and a Lorentzian squared line shape was used for the transverse scans.  Together these graphs show neither the FWHM nor the peak position changes significantly through T$_C$.
Measurements at (1/2 1/2 2.7) (not shown) indicate no change in the background through T$_C$.
Thus, there is either a reduction in the ordered magnetic moment or a reduction in the fraction of the sample that is magnetically ordered that is coincident with the onset of superconductivity in BaFe$_{1.92}$Co$_{0.08}$As$_2$.
A somewhat larger reduction in magnetic Bragg peak intensity has also been seen by Pratt, \textit{et al.}\cite{pratt} for a sample with a slightly higher Co-doping and consequently a higher T$_C$ and lower T$_N$.  Together these results show that competition between magnetism and superconductivity is robust in the underdoped region of these materials in samples with different T$_C$ and synthesized by different groups.

Naturally the question arises: What happens to the spectral weight associated with the reduced Bragg peak intensity below T$_C$?  As we discuss below, the spin excitations may provide the answer.  Fig. 2(a) shows a constant \textbf{Q} scan at the (1/2 1/2 -1) wave vector at 1.6 ($T<T_C$) and 20 K ($T>T_C$).  For comparison, we show the background estimated from combining constant-E scans with the constant-Q scan at (0.65 0.65 -1).  This scattering has only negligible temperature dependence suggesting that the background is predominately instrumental in origin.  The data at 1.6 K show an additional intensity at an energy of $\sim$5 meV.   Figures 2(c) and (d) show the in-plane and out-of-plane wave vector dependence of the spin excitations above and below T$_C$ near (1/2 1/2 -1).  Unlike  optimally doped samples\cite{lumsden_co} the intensity is strongly peaked both in-plane and out-of-plane indicating 3d spin correlations.  
Figure 2(b) shows the evolution of the extra intensity as a function of temperature.  A powerlaw fit yields a T$_C$ of 11(1) K in agreement with the bulk measurements on these samples demonstrating a strong coupling between the appearance of the resonance and superconductivity.

\begin{figure}
\centering\includegraphics[width=1.0\columnwidth]{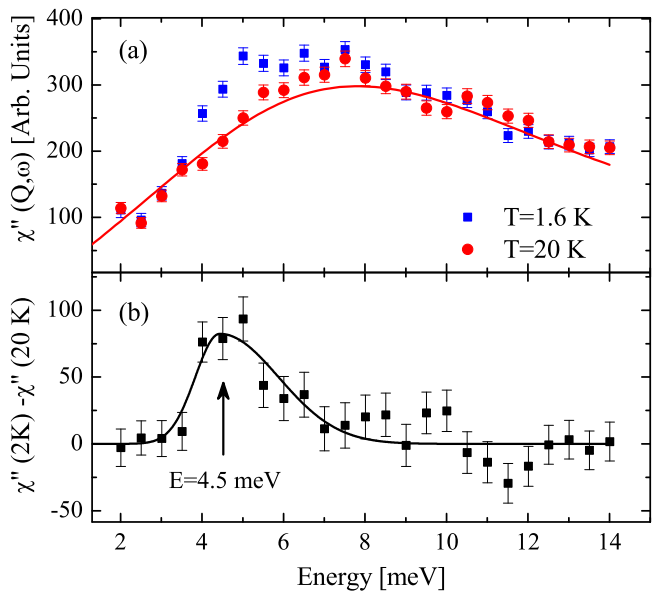}
\caption{\label{fig3} (a) $\chi^{''}$(\textbf{Q},$\omega$) for \textbf{Q}=(1/2 1/2 -1) at 1.6 and 20 K.  The line is a fit to $\chi{''}$ at 20 K as described in the text.  (b) The difference between $\chi^{''}$(\textbf{Q},$\omega$) at 1.6 and 20 K showing an additional component to the spin dynamics which appears below T$_C$.}
\end{figure}

In the optimally doped samples, the intensity gained at the resonance energy for T$<$T$_C$ is compensated by the opening of a spin gap at lower energy \cite{christianson2,lumsden_co,chi_ni}.  To search for such a gap in this case, thermal population effects must be removed.  To this end, the dynamic susceptibility ($\chi^{''}$(\textbf{Q},$\omega)$) has been derived from the data in Fig. 2(a) by removing the background scattering, correcting for thermal population, and applying a correction for higher order contamination in the beam monitor. The resulting $\chi^{''}$(\textbf{Q},$\omega)$ at 1.6 and 20 K is shown in fig. 3(a).   This shows that a spin gap is already present at 20 K and that an additional gap does not open below T$_C$ for energies as low as 2 meV.  In this case, a likely origin for the additional spectral weight required for the change in $\chi^{''}$(\textbf{Q},$\omega)$ is the loss of spectral weight at the magnetic Bragg peak positions. This data appears to be in contrast to the conclusions of \cite{pratt} which claim gapless spin excitations below T$_N$ with a gap opening below T$_C$.  This difference could be attributed to the slightly different Co-doping.  However, we note that our measurements extend over a wider energy range and were performed with fixed E$_f$ which allows for determination of S(\textbf{Q},$\omega$) with fewer corrections to the data.

\begin{figure}
\centering\includegraphics[width=1.0\columnwidth]{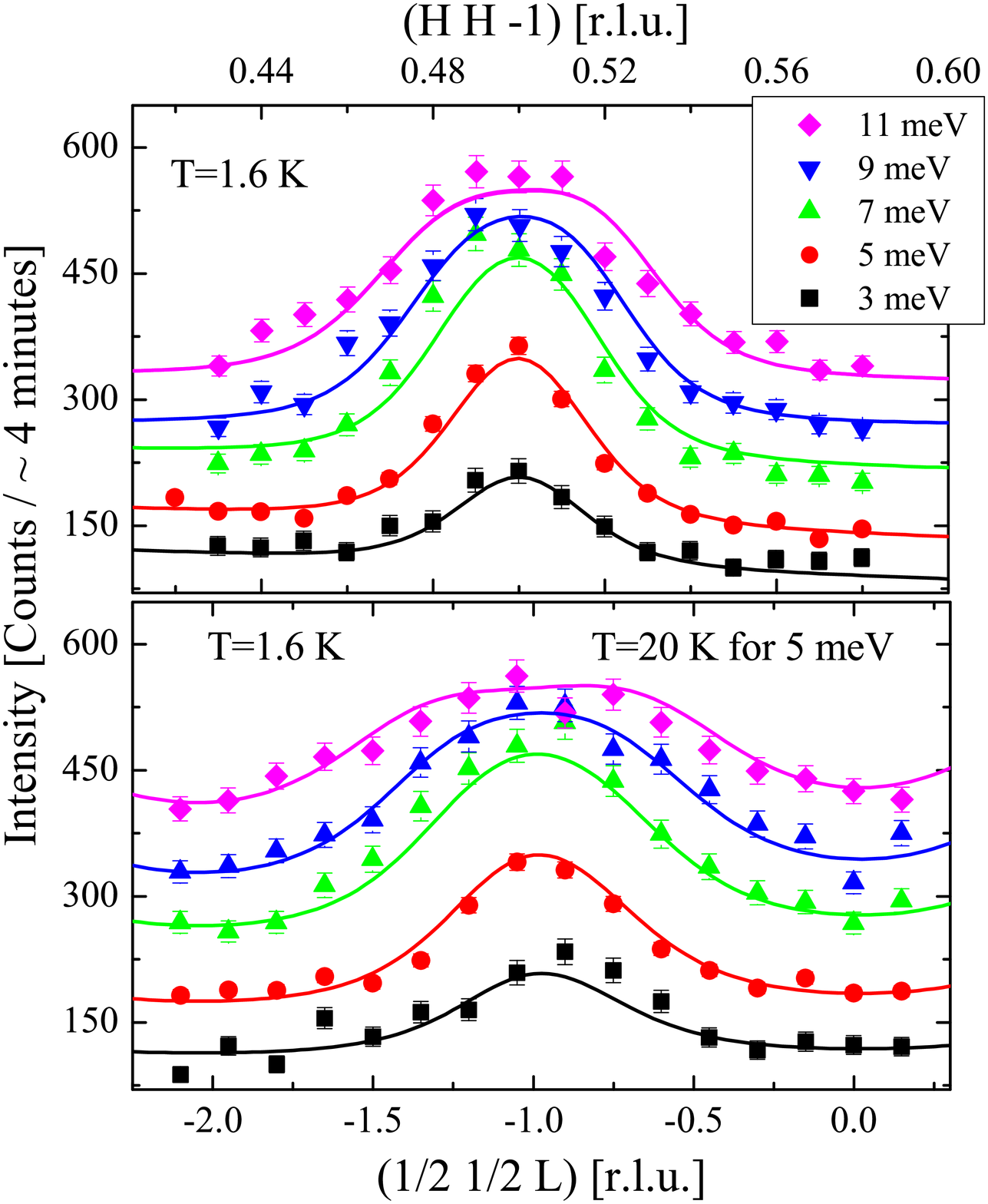}
\caption{\label{fig4} (a) Constant-E scans along (H H -1). (b) Constant-E scans along (0.5 0.5 L). All data is measured at 1.6K except E=5 meV, measured at 20 K to prevent the introduction of spin resonance scattering.  The solid lines represent fits to the model with a single amplitude as described in the text .}
\end{figure}

Figure 3(b) shows the difference between the dynamic susceptibility, $\chi{''}$($\textbf{Q}$=(1/2 1/2 -1),$\omega$) at 1.6 and 20 K.   This emphasizes the appearance of the resonant intensity appearing below T$_C$ with a maximum at E=4.5(0.5) meV or 4.7 k$_B$T$_C$. This is consistent with the energy of the spin resonance observed for other Fe-based superconductors\cite{christianson2,lumsden_co,chi_ni} indicating universality across the phase diagram.  In contrast to the data reported for an optimally Co-doped sample \cite{lumsden_co}, the spin resonance intensity varies strongly with L and is only observed for L near the antiferromagnetic zone center as shown in Fig. 2(c).

To explore the spin excitations in more detail, a series of constant-E scans along (H H -1) and (1/2 1/2 L) are shown in Fig. 4.  These measurements show data peaked in both H and L for energies up to 11 meV indicating anisotropic 3d interactions as in the parent compound \cite{matan}. To quantitatively analyze the spin wave dispersion, we consider a Heisenberg model with in-plane near-neighbor coupling constants $J_{1a}$ and $J_{1b}$, next near-neighbor coupling constant $J_2$, a c-axis coupling constant, $J_c$ and anisotropy, D as described previously \cite{ewings,matan,popnote}. The dispersion for this Hamiltonian is $\omega_q=\sqrt{A_q^2-B_q^2}$ where $A_q=2S[(J_{1b}\cos(\pi(H-K))-1)+J_{1a}+J_{c}+2J_2+D]$, $B_q=2S[J_{1a}\cos(\pi(H+K))+2J_2\cos(\pi(H+K)\cos(\pi(H-K)))+J_c \cos(\pi L)]$.  To analyze the data, we convolved the following S(\textbf{Q},$\omega$) (\emph{e.g}. \cite{ewings}) with the instrumental resolution function:
\begin{equation}
S(\bf{Q},\omega)\propto\frac{\it{A_q}-\it{B_q}}{\omega_q}\frac{4}{\pi}\frac{\Gamma \omega \omega_q}{\left(\omega^2-\omega_q^2\right)^2+4(\Gamma \omega)^2}
\end{equation}
Fits were performed to constant-E scans measured at 1.6 K except for E=5 meV where data taken from 20 K was analyzed to prevent the inclusion of resonance intensity.  Excluding the 5 meV data from the fits does not significantly alter the results.
The data was corrected for the Bose factor and the contamination of the beam monitor by higher order scattering from the monochromator.  The model describes the data well with the inclusion of a large damping parameter, $\Gamma$=6.2 meV.  Similar broad scattering was observed in measurements on BaFe$_2$As$_2$ \cite{matan}.  To properly describe a heavily damped spin excitation a damped harmonic oscillator, as opposed to a Lorentzian, is essential to prevent divergence of $\chi{''}$ at small energies.  The solid lines shown in Fig. 4 as well as the solid line through the 20 K constant-Q scan in Fig. 3(a) are the results of the fits with a common amplitude.
Measurements in the (HHL) scattering plane at these energies are insensitive to $J_{1b}$ and only depend on $J_{1a}$ and $J_2$ via $J_{1a}$+2$J_{2}$.  This was confirmed by sconsidering multiple values of J$_{1b}$ and ratios of J$_{1a}$/2J$_{2}$ all yielding identical results.  The best fit values are $S(J_{1a}+2J_2)$=32 meV, $SJ_{c}$=0.34 meV, and $SD$=0.25 meV yielding a gap energy of 8 meV.  For comparison with previous measurements on parent compounds, we can use these values to calculate the spin wave velocities resulting in values of $v_{\parallel}$=180 meV{\AA} and $v_{\perp}$=43 meV{\AA} yielding an anisotropy ratio of 4.2.  This value is consistent with measurements on BaFe$_2$As$_2$ \cite{matan}indicating no change in anisotropy.  This is in contrast with the optimally doped measurements where the excitations were found to be much more two-dimensional \cite{lumsden_co}.  This suggests that the presence of the structural phase transition may be responsible for the three dimensionality in these systems, a conclusion which is consistent with recent ARPES measurements \cite{arpes}.

In summary, neutron scattering measurements on single crystals of BaFe$_{1.92}$Co$_{0.08}$As$_2$ indicate spin waves which exhibit a gap of 8 meV and anisotropic 3d interactions.  On cooling below T$_C$, a reduction in the magnetic Bragg peak intensity of 6 $\%$ is observed with the simultaneous appearance of a resonance at the same wave vector with a characteristic energy of 4.5 k$_B$T$_C$.  In addition, there is no evidence of an additional spin gap developing below T$_C$.  Taken together these conclusions support a picture where spectral weight is transferred from ordered magnetic moments to spin fluctuations in BaFe$_{1.92}$Co$_{0.08}$As$_2$ in contrast to optimally doped samples \cite{christianson2,lumsden_co,chi_ni} with no long range magnetic order where a gap in the excitation spectrum opens contributing the spectral weight required to produce a resonance.

We acknowledge useful discussions with A. Zheludev, D. Singh,  H. Mook, T. Egami, T. Maier, and D. Scalapino. This work was supported by the Scientific User Facilities Division and the Division of Materials Sciences and Engineering, Office of Basic Energy Sciences, DOE.


\begin{thebibliography}{99}
\bibitem{popnote} Since the low temperature structure is orthorhombic, two domains of equal population were included in the analysis.
\end{thebibliography}
\end{document}